\documentclass[prb,twocolumn,showpacs,superscriptaddress,amsmath,amssymb]{revtex4}
\usepackage{graphicx,color}
\usepackage{dcolumn}
\usepackage{bm}
\begin{document}
\title{Properties of atomic intercalated boron nitride $K_{4}$ type crystals}

\author{Masahiro Itoh}
\affiliation{Institute of Multidisciplinary Research for Advanced Materials, Tohoku University, Aoba-ku, Sendai 980-8577, Japan}
\author{Seiichi Takami}
\affiliation{Institute of Multidisciplinary Research for Advanced Materials, Tohoku University, Aoba-ku, Sendai 980-8577, Japan}
\author{Yoshiyuki Kawazoe}
\affiliation{Institute for Materials Research, Tohoku University, Aoba-ku, Sendai 980-8577, Japan}
\author{Tadafumi Adschiri}
\affiliation{Institute of Multidisciplinary Research for Advanced Materials, Tohoku University, Aoba-ku, Sendai 980-8577, Japan}
\affiliation{Advanced Institute for Materials Research, WPI, Tohoku University, Aoba-ku, Sendai 980-8577, Japan}
\date{\today}

\begin{abstract}
The stability of atomic intercalated boron nitride $K_{4}$ crystal structures,
XBN (X=H, Li, Be, B, C, N, O, F, Na, Mg, Al, Si, P, S, Cl, K, Ca, Ga, Ge, As, Se, Br, Rb or Sr)
is evaluated by the geometric optimization and frozen phonon calculations based on the first principles calculations.
NaBN, MgBN, GaBN, FBN and ClBN are found to be stable.
NaBN, GaBN, FBN and ClBN are metallic, whereas MgBN is semiconducting.
\end{abstract}

\maketitle

\section{Introduction}

After the study of crystal structure by Wells\cite{Wells},
Sunada discovered the 3-D periodic net structure, called a $K_{4}$ crystal, from the graph theory\cite{Sunada}.
Interestingly, the $K_{4}$ crystal is believed to have three bonds per atom, similar to graphite.
Therefore, the crystal structure can be believed to have features intermediate between those of diamond and graphite.
However, its existence and properties were not investigated in detail.

Subsequently, Rignanese and Charlier\cite{RC-PRB-2008}, and Itoh \textit{et al.}\cite{IKNSKA-PRL-2009}
investigated its stability and electronic properties by first principles calculations based on the density functional theory\cite{HK}. 
However, phonon calculations\cite{Yao, Carbide-K4-ITKA} showed that this possibility is very small.
Although the $K_{4}$ type crystal structure is thermally unstable,
its crystal frame is found to be thermally stable when distorted by doping impurities such as Na or Mg\cite{Carbide-K4-ITKA}.

Can other $K_{4}$ type crystal structure be found in nature ?
Mathematically, the $K_{4}$ crystal structure can be constructed
by two different types of elements, as shown by Kotani \textit{et al}\cite{Kotani}.
The BN system is the simplest example.

Crystalline BN is a fascinating material because of its high thermal conductivity,
extreme hardness, high melting temperature and wide band gap.
In contrast to diamond, BN does not react with iron and steel,
making it a promising candidate for coating high duty toolsets.

As in the recent situation regarding the carbon and atomic intercalated carbon systems,
the stability and properties $K_{4}$ crystal for boron nitride and atomic intercalated boron nitride systems have not been investigated.
Such investigation will provide valuable knowledge about the system stability and properties.

This study examines the structural stability as well as several properties
of BN and XBN (X = H, Li, Be, B, C, N, O, F, Na, Mg, Al, Si, P, S, Cl, K, Ca, Ga, Ge, As, Se, Br, Rb or Sr) crystals
with a $K_{4}$ type BN frame by first principles calculations.

\section{Computational Methods}
First principles calculations based on density functional theory\cite{HK, KS}
were performed for BN and XBN crystals with $K_4$ type of frame
using the Vienna Ab-initio Simulation Package (VASP)\cite{Kresse}.
Local density approximation (LDA)\cite{PZ, CA} is used for the exchange-correlation energy functional.
In these calculations, all the crystals are considered as spin-un polarized systems.

Computational costs were reduced by employing the projector-augmented wave method\cite{PAW}
to approximate nucleus, inner core electrons and valence electrons in each atom of the crystal.
For H, Li, Be, B, C, N, O, F, Na, Mg, Al, Si, P, S, Cl, K, Ca, Ga, Ge, As, Se, Br, Rb and Sr,
valence electrons of 1$s^{1}$,
2$s^{1}$, 2$s^{2}$, 2$s^{2}$2$p^{1}$, 2$s^{2}$2$p^{2}$, 2$s^{2}$2$p^{3}$,
2$s^{2}$2$p^{4}$, 2$s^{2}$2$p^{5}$,
3$s^{1}$, 3$s^{2}$, 3$s^{2}$3$p^{1}$, 3$s^{2}$3$p^{2}$, 3$s^{2}$3$p^{3}$,
3$s^{2}$3$p^{4}$, 3$s^{2}$3$p^{5}$,
3$s^{2}$3$p^{6}$4$s^{1}$, 3$s^{2}$3$p^{6}$4$s^{2}$, 4$s^{2}$4$p^{1}$, 4$s^{2}$4$p^{2}$, 4$s^{2}$4$p^{3}$,
4$s^{2}$4$p^{4}$, 4$s^{2}$4$p^{5}$,
4$s^{2}$4$p^{6}$5$s^{1}$ and 4$s^{2}$4$p^{6}$5$s^{2}$ are considered, respectively.

The following procedure is adopted to
evaluate the stability of the XBN $K_{4}$ crystal structure.

(1) An original primitive unit cell for the BN $K_{4}$ crystal is shown in Figure 1(a).
Four X atoms (X = H, Li, Be, B, C, N, O, F, Na, Mg, Al, Si, P, S, Cl, K, Ca, Ga, Ge, As, Se, Br, Rb or Sr) are allocated to reduced coordinates
($\frac{1}{8}$, $\frac{1}{8}$, $\frac{1}{8}$), ($\frac{7}{8}$, $\frac{5}{8}$, $\frac{3}{8}$),
($\frac{3}{8}$, $\frac{7}{8}$, $\frac{5}{8}$), ($\frac{5}{8}$, $\frac{3}{8}$, $\frac{7}{8}$) in the primitive unit cell.
The primitive unit cell of XBN with $P2_{1}3$ ($T^{4}$) symmetry is shown in Fig. 1(b).
The symmetric plane of the atomic intercalated system is shown in Fig. 1(c).
As Fig. 1(d) shows, three adjacent atoms (red) are aligned on a straight line at equal distances
(B-X and X-N (B$_{A}$-X and X-$N_{A}$ shown in Fig. 1(e))).

(2) The binding energy vs volume curve is evaluated and fitted by Murnaghan's equation of state\cite{Murnaghan}.
For these calculations, Brillouin zone integration is performed
for 8$\times$8$\times$8 \textit{\textbf{k}}-point meshes generated by the Monkhorst-Pack scheme.
The method of residual minimization/direct inversion in the iterative subspace
is used to accelerate the convergence of self-consistent total energy calculations.
The convergence criterion was set to be within 1 $\times 10^{-8}$ eV/formula unit cell.
The cut-off energy for the plane-wave expansion of valence electrons for the primitive unit cell was determined,
so as the number of plane waves to be constant over the full range of lattice constant.
Around the minimum of binding energy vs. volume curve.
The cut off energies were set to be 500 eV for BN and XBN crystals.

(3) The unit cells obtained from step (2) are optimized with freedom of the atomic configuration under the symmetry constraint.

(4) The unit cells obtained from step (3) are optimized without any constraint in the crystal structure.
The convergence criterion for step (2-4) was set to be within 1 $\times 10^{-7}$ eV/\AA\ unit cell.

(5) Frozen phonon calculations are perfomed using the FROPHO code\cite{FROPHO}
which is based on the Parlinski-Li-Kawazoe method\cite{Parlinski-Li-Kawazoe}.
To obtain the force constants for the phonon calculations, the atomic displacements are set to be 0.01 \AA\ .
The Born - von Karman boundary condition is applied for each obtained primitive unit cell with multiplication
after step (3) and (4) for the phonon calculation.
In these calculations, the primitive cells are multiplied by two for each direction of the unit vector.

\section{Results and Discussions}

In the present study, most of the crystal structures with a $K_{4}$ type frame considered for BN and XBN systems
maintain the $K_{4}$ type crystal frame composed of B and N atoms with distortion through the entire process of geometrical optimization.
However, SiBN, GeBN and RbBN are transformed to stacked BN decagonal ring-layer crystal structures with Si, Ge
and B-N-B-N-B-N strip crystal structures with Rb, respectively, as shown in Fig. 2. 
In all other cases, the crystal symmetry is maintained within an accuracy of less than 0.001\AA\ .
The lattice constant ($a$), volume at lattice constant ($V_{0}$), cohesive energy per atom ($E_{coh}$) 
and bulk modulus at $V_{0}$ ($B_{0}$) are evaluated for the fully optimized structure.
In this study, $E_{coh}$ for XBN, $E_{coh, XBN}$, is defined as
\begin{widetext}
\begin{equation*}
E_{coh, XBN}\equiv\frac{-E_{XBN}+N_{X atom}\times E_{X atom}+N_{B atom/K_{4}}\times E_{B atom}+N_{N atom/K_{4}}\times E_{N atom}}
{N_{X atom}+N_{B atom/K_{4}}+N_{N atom/K_{4}}}.
\end{equation*}
\end{widetext}
Here, $E_{A}$ and $N_{A}$ are the total energy and number of atoms for A, respectively.
To obtain these values, Murnaghan's equation of state is used to fit the binding energy vs. volume curve.
Parameters in the equation are determined using the least squares method in the range of about $0.8V_{0} < V < 1.2V_{0}$,
where the root mean square is set to be less than 5.0 meV/atom.
The obtained values of $a$, $V_{0}$, $E_{coh}$ and $B_{0}$ for the crystals are shown in Fig. 3(a).

In XBN, the values of $a$ and $V_{0}$ increase with increasing atomic number of X.
Generally, the values of $a$ for XBN crystals are larger than that of boron nitride $K_{4}$.
The boron nitride $K_{4}$ frames are mostly expanded by intercalating X to the boron nitride system.

However, the values of $V_{0}$ for X = H, Li, .., S, are smaller than that of boron nitride $K_{4}$ whereas those for X = Cl, .., Sr are larger.
This is because boron nitride $K_{4}$ has sufficiently wide vacant spaces that are filled by X atoms,
and the spaces occupied by X atoms becomes larger with increasing atomic number of X.

On the other hand, $E_{coh}$ decreases with increasing period of X.
$E_{coh}$ is strongly negatively correlated with $a$ and $V_{0}$,
as shown in the correlation coefficients: $r_{a-E_{coh}}$ = -0.770 and $r_{V_{0}-E_{coh}}$ = -0.769.
The $E_{coh}$ values of XBN are smaller than boron nitride $K_{4}$ values.
These results suggest that the X-B and X-N bonds in XBN crystals are weaker than the B-N bonds in the boron nitride $K_{4}$ crystal.
For each period of X in XBN, $E_{coh}$ generally shows the largest values for IIIB(IVB) elements of X and the smallest values for IA or VIIB.
The maximum energy gain for X = IIIB(IVB) can be explained by completion or over completion
of the electronic shell in the X, B and N atoms composing the $K_{4}$ type frame of XBN crystals.
Subsequently, no correlation appears between $E_{coh}$ and the dynamical stability.

Like $E_{coh}$, $B_{0}$ also decreases with increasing period of X.
$B_{0}$ and $E_{coh}$ are strongly correlated,
as shown in the correlation coefficient: $r_{E_{coh}-B_{0}}$ = 0.837.
$B_{0}$ values in XBN are generally smaller than that of BN.

The trend of each physical value is qualitatively the same as in XC$_{2}$ systems\cite{Carbide-K4-ITKA}.
However, the correlation coefficients for $a$, $V_{0}$, $E_{coh}$ and $B_{0}$ in XBN show
slightly smaller absolute values than those of XC$_{2}$ systems:
$r_{a-E_{coh}}$ = -0.792, $r_{V_{0}-E_{coh}}$ = -0.796 and $r_{E_{coh}-B_{0}}$ = 0.848.
This suggests that the properties of XBN systems can be understood by our knowledge of XC$_{2}$ systems.

To better understand the correlation between stability and structure,
the nearest-neighbour distances and angles between different atoms specified in Figure. 1(e) are investigated
[$d$ B-N,  $d$ X-B$_A$, $d$ X-B$_B$, $d$ X-B$_C$, $\overline{d}$ X-B (the average of all $d$ X-B values),
$d$ X-N$_A$, $d$ X-N$_B$, $d$ X-N$_C$, $\overline{d}$ X-N (the average of all $d$ X-N values) and $d$ X-X,
and the angles $\angle$ X-B$_{A}$-N$_{C}$, $\angle$ X-N$_{A}$-B$_{C}$ and dihedral angle $\angle$ B-N-B-N].

Surprisingly, the relative position of X in the crystal is maintained, yielding $r_{a-d X-X}$ = 0.999.
That is, $d$ X-X determines the volume of the crystal.
Therefore, $d$ X-X can be considered as a standard distance in XBN systems.
To emphasise the similarity of the structures,
the various distance ratios of the nearest-neighbour distances to the $d$ X-X are shown in Figs. 4(a) and (b).

As Figure 3(b) shows, $d$ B-N, $\overline{d}$ X-B, $\overline{d}$ X-N and $d$ X-X,
like $a$ and $V_{0}$, show certain trends with increasing atomic number of X.
$d$ B-N, $\overline{d}$ X-B and $\overline{d}$ X-N are very strongly correlated with $a$,
as shown in the correlation coefficients: $r_{a-d B-N}$ = 0.842, $r_{a-\overline{d} X-B}$ = 0.973 and $r_{a-\overline{d} X-N}$ = 0.990.
Based on the values of $d$ X-X, the values of $\overline{d}$ X-B and $\overline{d}$ X-N are possibly determined.
$d$ B-N seems to be determined indirectly by $\overline{d}$ X-B, $\overline{d}$ X-N and $d$ X-X.
On the other hand, these distances have weaker correlations with $E_{coh}$:
$r_{E_{coh}-d B-N}$ = -0.592, $r_{E_{coh}-\overline{d} X-B}$ = -0.699, and $r_{E_{coh}-\overline{d} X-N}$ = -0.723.
Also note that $\overline{d}$ X-B and $\overline{d}$ X-N, like $a$ and unlike $d$ B-N, show strong correlation with $E_{coh}$.

Here, the unequal atomic distances between X and B atoms and X and N atoms
($d$ X-B$_{A}$, $d$ X-B$_{B}$, $d$ X-B$_{C}$, $d$ X-N$_{A}$, $d$ X-N$_{B}$ and $d$ X-N$_{C}$) should be noted.
It is apparent that the correlations of $\overline{d}$ X-B and $\overline{d}$ X-N with $a$ and $E_{coh}$
result from the correlation of $d$ X-B$_{A}$, $d$ X-B$_{B}$, $d$ X-B$_{C}$, $d$ X-N$_{A}$, $d$ X-N$_{B}$ and $d$ X-N$_{C}$
with $a$ and $E_{coh}$ as shown in the correlation coefficients:
$r_{a-d X-B_{A}}$ = 0.543, $r_{a-d X-B_{B}}$ = -0.042, $r_{a-d X-B_{C}}$ = 0.927,
$r_{a-d X-N_{A}}$ = 0.812, $r_{a-d X-N_{B}}$ = 0.338, $r_{a-d X-N_{C}}$ = 0.864,
$r_{E_{coh}-d X-B_{A}}$ = -0.436, $r_{E_{coh}-d X-B_{B}}$ = 0.161, $r_{E_{coh}-d X-B_{C}}$ = -0.781,
$r_{E_{coh}-d X-N_{A}}$ = -0.808, $r_{E_{coh}-d X-N_{B}}$ = -0.073, and $r_{E_{coh}-d X-N_{C}}$ = -0.629.
These values indicate that $d$ X-N$_{A}$, $d$ X-B$_{C}$ and $d$ X-N$_{C}$ ($d$ X-B$_{A}$)
are mainly determined by the volume and the energy gain. 

For each group of X, $d$ X-B$_{A}$, $d$ X-B$_{B}$, $d$ X-B$_{C}$,
$d$ X-N$_{A}$, $d$ X-N$_{B}$ and $d$ X-N$_{C}$ increase with increasing atomic number of X.
Here, for each period of X, $d$ X-B$_{A}$($d$ X-N$_{A}$) increases(decreases) up to X = IVB or VB and then decreases(increases).
Apparently, this trend is determined by the number of valence electrons of X.
However, an opposite trend is shown in a system with X of second period.
This can be attributed to the similarity between elements of $K_{4}$ type frame and X.
Here, we should note the strong (weak) negative correlation
between $d$ X-B$_{A}$ and $d$ X-B$_{B}$ ($d$ X-N$_{A}$ and $d$ X-N$_{B}$):
$r_{d X-B_{A}-d X-B_{B}}$ = -0.818 ($r_{d X-N_{A}-d X-N_{B}}$ = -0.134).

As Table III in the Appendix shows, for each third and forth period with increasing atomic number of X, 
the dihedral angle $\angle$ B-N-B-N decreases up to X = IVB and then increases.
For each period, the dihedral angle becomes minimum at X = IVB, which breaks the $K_{4}$ type frame structure.
For the values of $d$ X-B$_A$, $d$ X-B$_B$, $d$ X-N$_A$ and $d$ X-N$_B$
which have larger amplitudes compared with $d$ X-B$_C$ and $d$ X-N$_C$,
the correlation with $\angle$ B-N-B-N are investigated.
Direct correlation between $d$ X-B$_{A}$, $d$ X-B$_B$, $d$ X-N$_{A}$, $d$ X-N$_{B}$ and $\angle$ B-N-B-N:
$r_{{d X-B_{A}}-{\angle B-N-B-N}} = -0.798$, $r_{{d X-B_{B}}-{\angle B-N-B-N}} = -0.821$,
$r_{{d X-N_{A}}-{\angle B-N-B-N}} = 0.196$, and $r_{{d X-N_{B}}-{\angle B-N-B-N}} = -0.145$
shows much preference of $d$ X-B$_{A}$ and $d$ X-B$_{B}$ than $d$ X-N$_{A}$ and $d$ X-N$_{B}$.
X dependence of the angle $\angle$ X-B$_{A}$-N$_{C}$ and $\angle$ X-N$_{A}$-B$_{C}$
have stronger correlation with dihedral angle $\angle$ B-N-B-N:
$r_{{\angle X-B_{A}-N_{C}}-{\angle B-N-B-N}}$ = 0.888 and $r_{{\angle X-N_{A}-B_{C}}-{\angle B-N-B-N}}$ = -0.662.
$\angle$ X-B$_{A}$-N$_{C}$($\angle$ X-N$_{A}$-B$_{C}$) depends strongly on that of $d$ X-B$_{A}$ and $d$ X-N$_{A}$:
$r_{{d X-{B_{A}}}-{\angle{X-B_{A}-N_{C}}}}$ = -0.964 and $r_{{d X-{N_{A}}}-{\angle{X-N_{A}-B_{C}}}}$ = -0.772.
These suggest that X dependence of $d$ X-N$_{A}$ is indirectly correlated to $\angle$ B-N-B-N,
through and $\angle$ X-N$_{A}$-B$_{C}$ while $d$ X-B$_{A}$ is directly correlated.
The values are determined by the number of outer-most valence electrons and the occupied space of X.
Weak correlation of $\angle$ B-N-B-N with $a$, $V_{0}$, $E_{coh}$ and $B_{0}$ should be also noted.

The distinct difference between X = IA and IIA and the other X atoms found in XC$_{2}$ systems
with X of the third, forth(fifth) periods in the periodic table\cite{Carbide-K4-ITKA} are not found in XBN systems.
Note the exceptional features of BeBN in the distance and angle trends.
Apparently, this results from the higher complexity of the bonds formed in XBN systems relative to those of XC$_{2}$ systems.
However, the relative atomic configurations of XBN systems can be considered to have some deviation from XC$_{2}$
on the basis of a comparison between the normalized distances shown in Fig. 2(c) in ref.\cite{Carbide-K4-ITKA} and Figs. 4(a) and (b).

In Figure 5, the phonon density of states(DOS) without step (4) for BN and XBN
(X = H, Li, Be, B, C, N, O, F, Na, Mg, Al, Si, P, S, Cl, K, Ca, Ga, Ge, As, Se, Br, Rb or Sr), with $K_{4}$ type frame of B and N, are depicted.
In general, the bandwidth of XBN depends on the period of X, which is the narrowest in third period.
Thus, the inter-atomic distances approximately determine the vibrational frequency range in these crystals.
As shown in the figure, imaginary frequency modes rarely appear in NaBN, MgBN, GaBN, FBN and ClBN.

For the obtained structures in BN, NaBN, MgBN, GaBN, FBN and ClBN after step (4),
the DOS and dispersion relationship are also evaluated.
As Fig. 6 shows, the phonon DOS for the fully optimized structures of other XBN crystals show
almost the same shapes as those of the optimized structures from step (3).
Therefore, as in the previous study\cite{Carbide-K4-ITKA}, the discussion in the previous paragraph is expected to be significant,
and omitting step (4) is effective for reducing the computational cost.
However, exceptional cases such as SiBN, GeBN and RbBN exist, as shown in Fig. 2.

Figure 6 shows that boron nitride $K_{4}$ has imaginary frequency modes at a wide range of $\textbf{\textit{k}}$ points.
This suggests a short life-span of the boron nitride $K_{4}$ in nature, similar to carbon $K_{4}$\cite{Yao, Carbide-K4-ITKA}.
On the other hand, although NaBN, MgBN, GaBN, FBN and ClBN have imaginary frequencies in the acoustic modes around the $\Gamma$ point,
no imaginary modes appear at the other $\textbf{\textit{k}}$ points.
This suggests that the structures are stable, although
they break against phonon vibration with a long wavelength limit.
From previous calculations for XC$_{2}$ and SrSi$_{2}$\cite{Carbide-K4-ITKA},
the stable existence of NaBN, MgBN, GaBN, FBN and ClBN $K_{4}$ type crystals in nature is expected
although maximum values of $E_{coh}$ are not realized for the considered XBN crystals.

From the maximum frequencies obtained for BN diamond (1300 cm$^{-1}$\cite{Tohei}) and graphite (1600 cm$^{-1}$\cite{Tohei}),
the orders of the B-N chemical bonds in XBN are considered as at most 1.
Therefore, the orders are expected to be less than 1 for NaBN, MgBN, GaBN, FBN and ClBN crystals.
The feature of the chemical bonds and the stability of $K_{4}$ systems are analyzed by examining the valence charge distribution.

Figure 7 shows the various type of distribution in the primitive unit cells
for fully optimized BN, NaBN, MgBN, GaBN, FBN and ClBN crystals with a distorted $K_4$ type BN frame.
Fig. 7(a) details the planes selected to show the distributions of the quantities noted in Figs. 7(c), (e) and (g).
In Figs. 7(c) and (e), the high and low populations are neglected for better comparison.

Figures 7(b) and (c) show the isosurface and contour of charge density.
Apparently the valence charge is locally distributed along the lines between adjacent B and N atoms for the considered systems.
The charge is localized more to the N atoms than B atoms because of the electronegativity difference.
These features are also indicated in the zinc blende BN crystals\cite{BN}.
Although significant charge accumulation is shown for X in FBN and ClBN,
the difference in the charge distribution between BN and XBN is not so large in the BN $K_{4}$ frame.

Figures 7(d) and (e) show the differences between the valence charge density of the crystals (ex. $\rho_{NaBN}(\textbf{r})$)
and their separated components: the induced impurities (ex. $\rho_{Na}(\textbf{r})$) and the frame (ex. $\rho_{BN}(\textbf{r})$).
Let us examine the distribution around the X atoms.
Significant excess charge accumulates at the X atom only in FBN and ClBN and not in NaBN, MgBN and GaBN.
Between adjacent X atoms the charge excess is low.
The higher excess and depressed charge for FBN compared to ClBN seem to be related to the wider phonon frequency range.
From this example, it can be expected that a higher ionicity of X makes the system more robust.

Charge excess regions appear on the line between adjacent X and B$_{A}$ atoms,
and the region is closer to the X atoms for systems with larger electronegativity of X: FBN, ClBN, GaBN, MgBN and NaBN.
Further, as shown in these figures, significant excess charge appears
above the opposite side of the B$_{A}$ atom in the pseudoplane defined
by the B atom and three adjacent nearest-neighbour N atoms in NaBN, MgBN, GaBN, FBN and ClBN.
The resulting significant charge transfer causes ionic bonds between X and B$_{A}$.
For these cases, a region of depressed charge exists between the X and BN $K_{4}$ type frame.
The charge differences between the X and N$_{A}$ atoms
do not show much excess for NaBN, MgBN and GaBN, and are depressed for FBN and ClBN.

The charge accumulations in the region of the $\sigma$-type orbitals between adjacent B and N atoms are
depressed for NaBN, MgBN and GaBN but in excess around the region close to the N atom for FBN and ClBN.
However, features of the $\pi$-type bonds between adjacent B and N atoms are strengthened
in NaBN, MgBN and GaBN, whereas these bonds are not found in FBN and ClBN.
A correlation exists between the values of the dihedral angle $\angle$ B-N-B-N and the $\pi$ bonds.

The nature of the bonds is also investigated using the electronic localization function(ELF).
Figs. 7(f) and (g) show the isosurface (0.8 [-]) and the contour.
The ELF distributions of BN, NaBN, MgBN, GaBN, FBN and ClBN have some similarities despite of being significantly different.
The ELF accumulates distinctively around for Ga, F and Cl atoms but not so much for the Na and Mg atoms in XBNs.
First, let us note the distribution between the X atom and adjacent B$_{A}$ and N$_{A}$ atoms.
A region of significant ELF concentration appears between adjacent X and B$_{A}$ atoms for NaBN, MgBN, GaBN, FBN and ClBN.
Between adjacent X and N$_{A}$ atoms,
a region of concentrated ELF appears more closer to N$_{A}$ in NaBN, MgBN and GaBN and more closer to X in FBN and ClBN.
Between adjacent B and N atoms, features of the weakened $\sigma$ and strengthened $\pi$ character appear for NaBN, MgBN and GaBN
as expected in XC$_{2}$ systems\cite{Carbide-K4-ITKA},
whereas only the $\sigma$ character appears for FBN and ClBN.

Valence electron donation from X to B$_{A}$ in NaBN, MgBN and GaBN and from B$_{A}$ to X in FBN and ClBN
result in different types of ionic bonds between them.
The difference seems to originate mainly from the number of outermost valence electrons and the order of electronegativity in X.
$d$ X-B$_{A}$ becomes larger for electron donation from X in NaBN, MgBN and GaBN.
However, $d$ X-B$_{A}$ of FBN and ClBN is smaller than the other compounds.
We should also note that $d$ X-B$_{A}$/$d$ X-X of FBN and ClBN is smaller than that of BN, as shown in Figs. 4(a) and (b).
The X = F and Cl forms an electronic shell,
which results to strengthen $\pi$ bonds and weaken $\sigma$ bonds between adjacent B and N atoms
through reducing $d$ X-B$_{A}$ and increace $\angle$ B-N-B-N.
As shown here, the bonds between X and N$_{A}$ atoms are different from those between X and B$_{A}$ atoms.
The greater number of electrons in Ga compared with NaBN and MgBN shortens the relative distance between the X and N$_{A}$ atoms.
This results in the formation of the smallest $\angle$ B-N-B-N
and the distinctive $\pi$ bonds feature between adjacent B and N atoms among NaBN, MgBN, GaBN, FBN and ClBN.
The nature of the bonds in those crystals are gradually different with the number of valence electrons in X.
Here, we should note again that the net bond orders of XBN are less than that of BN from the phonon calculations.

Let us discuss the reason for the instability and stability of the BN and XBN systems.
In the BN $K_{4}$ crystal, the incomplete $\pi$ bonds
make the structure thermally unstable as in the C $K_{4}$ crystal.
The thermal stability of NaBN and MgBN can be attributed to the same reason
as that of NaC$_{2}$ and MgC$_{2}$\cite{Carbide-K4-ITKA}.
Valence electrons are donated from Na or Mg to the BN $K_{4}$ type frame.
In the $K_{4}$ type frame, polarized bonds between adjacent B and N atoms
are changed to strengthen the $\pi$ bonds and weaken the $\sigma$-type bonds.
The crystal structures become stable for the formation of these $\pi$ bonds.
Unlike NaBN and MgBN, KBN and CaBN are thermally unstable.
The difference can also be attributed to the amount of space occupied by the dopant X in XBN.

The thermal stability of GaBN can be understood by comparing it with other systems.
The instability of the subsequent crystal AlBN is believed to originate from the stronger electronegativity of Al compared to Na and Mg.
In AlBN, $\angle$ B-N-B-N is very small for the strong repulsion of B$_{A}$ by adjacent X.
The loss of the bonding balance puts the structure on the saddle with highly degenerated negative curvature on the potential surface.
Ga is less electronegative than Al, which results in less B$_{A}$ repulsion.
The larger occupied space of Ga in GaBN compared with that of Al in AlBN
is attributed to the greater amount of electrons in the inner core of Ga than in Al.
These differences seem to retain the $K_{4}$ type frame structure in GaBN
and make the $\angle$ B-N-B-N value the same as that of MgBN.
The system seems to be stabilized by not only the $\pi$-type bonds between adjacent B and N atoms but also the Ga-N$_{A}$ bonds.
Note that the large $d$ X-B$_{A}$ value of GaBN compared with $d$ X-N$_{A}$
which is as same as that of $d$ X-C$_{A}$ of GaC$_{2}$\cite{Carbide-K4-ITKA}.

In FBN and ClBN, the F and Cl distinctively attract B$_{A}$ and repel N$_{A}$ atoms.
This is due to the strong electronegativity of X, which attracts the valence electrons of neighbour B$_{A}$,
this results in the closing of the X atom's electronic shell.
The F atom attracts electrons more than the Cl atom because of its larger electronegativity,
and this results in smaller $d$ X-B$_{A}$ and larger $\angle$ B-N-B-N in FBN.
Between adjacent B and N atoms, although $\pi$ bonds do not form, the $\sigma$ bonds are strengthened.
In FBN and ClBN, the X-B$_{A}$ bonds and $\sigma$ bonds between adjacent B and N atoms probably stabilize the crystal structures.

The stabilization of the atomic intercalated BN $K_{4}$ crystals can be explained by the delicate balance of the bonding.
Discriminative stability seems to be realized by the proper atomic radii of X,
although the interpretation of thermal stability of the crystal structure is more difficult than in the case of XC$_{2}$ systems.
Future accumulation of knowledge will elucidate the stability of these systems.

Figure 8 shows X-ray diffraction (XRD) patterns of the fully optimized crystal structure of the BN, NaBN, MgBN, GaBN, FBN and ClBN.
Monochromatic radiation with wave length 1.541 \AA\ is assumed in these calculations.
As shown in this figure, the compounds have specific XRD peaks at larger $2\theta$ than pure BN crystals.

Figure 9 shows the electron band structures, DOS, and local DOS of valence electrons
for the atoms composing the $K_{4}$ type frame and the intercalated atoms in the fully optimized
BN, NaBN, MgBN, GaBN, FBN and ClBN crystals.
From the bottom energy levels to the upper levels, the angular momentum for the electronic states changes successively 
in the order of \textit{s}, \textit{p} and \textit{d} character, although every component appears in most of the states.
From the bottom energy levels to those around the Fermi level, the band structures are similar in these crystals.
NaBN, GaBN, FBN and ClBN $K_{4}$ crystals are metallic, whereas MgBN is semiconducting.

\section{Conclusions and Summary}

The stability of atomic intercalated boron nitride $K_{4}$ type crystals,
XBN (X=H, Li, Be, B, C, N, O, F, Na, Mg, Al, Si, P, S, Cl, K, Ca, Ga, Ge, As, Se, Br, Rb and Sr),
are evaluated using geometric optimization and frozen phonon analysis based on first principles calculations.
The phonon calculations show that NaBN, MgBN, GaBN, FBN and ClBN are stable.
NaBN, GaBN, FBN and ClBN are metallic while MgBN is semi-conducting.
\section{Appendix}
The values presented in Figures 3 and 4 are listed in Table I, II, and III.
Distance between adjacent B$_A$, X and N$_A$ atoms for fully optimized BN, NaBN, MgBN, GaBN, FBN and ClBN are shown in Fig. 10.

\begin{center}
{\bf Acknowledgements}
\end{center}
The authors acknowledge the staff of the CCMS-IMR for allowing the use of the HITACHI SR11000 supercomputing facilities.
The authors acknowledge Professor Motoko Kotani, Dr Hisahi Naito and Professor Toshikazu Sunada
for sharing the manuscript of $K_{4}$ type crystal structure composed of two kinds of element prior to publication.
We acknowledge financial support from the CREST, MEXT and WPI.

\begin{figure*}
\begin{center}
\includegraphics[width=12cm]{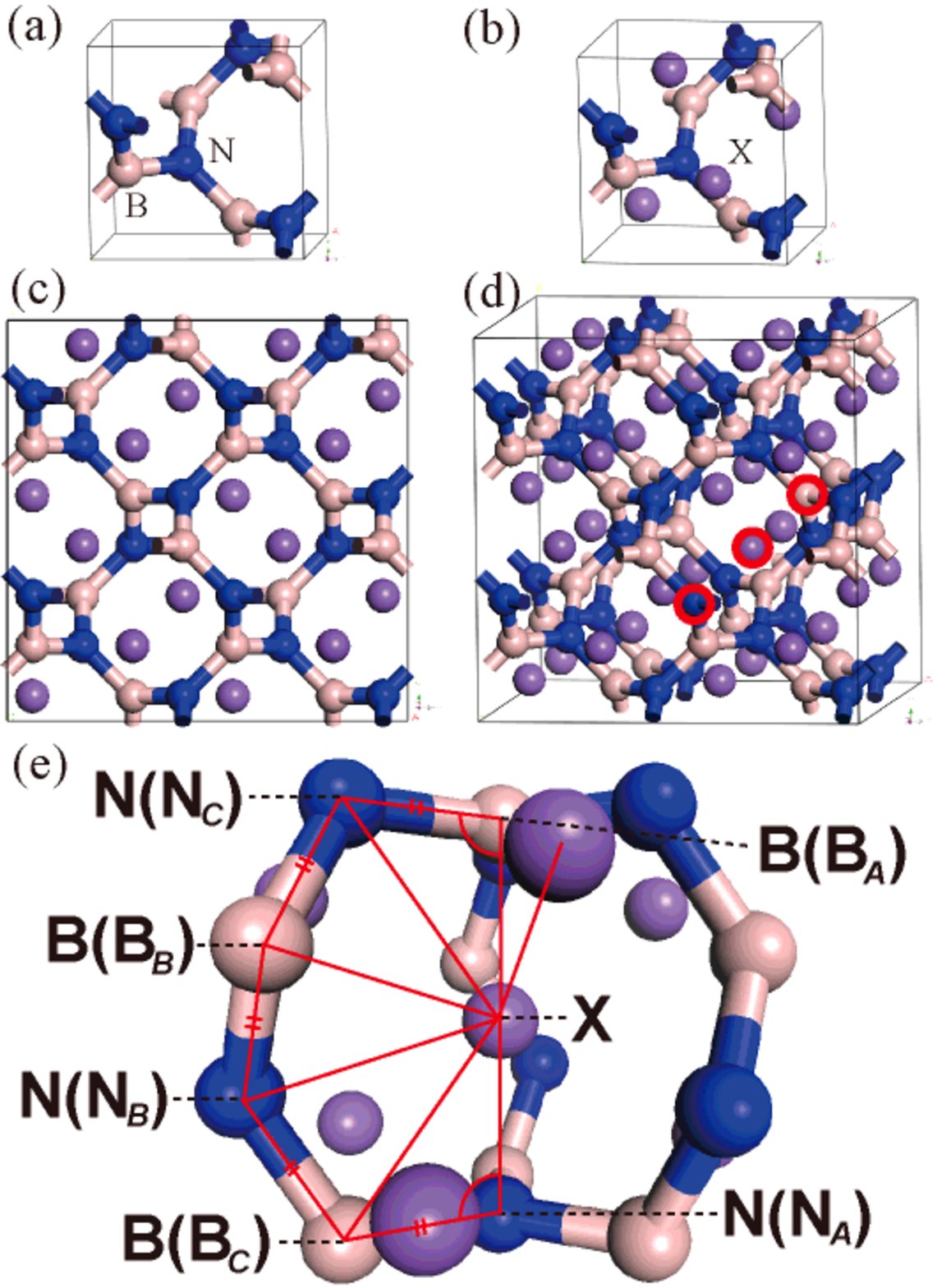}
\caption{\label{fig:K4-XBN-initial-structures-with-extraction}
(a) Primitive unit cell of the boron nitride (BN) $K_{4}$ crystal with $P4_{1}32(O^{7})$ symmetry.
(b) Initial structure of the primitive unit cell of the XBN crystal with $K_{4}$ frame of BN with $P2_{1}3(T^{4})$ symmetry.
(c) The 2$\times$2$\times$2 super-cell of (b) showing symmetric plane of the impurity intercalated system.
(d) The three adjacent atoms (B-X-N) aligned in a straight line at equal distances (B-X and X-N) in (c) are encircled with a thick red line.
(e) One of the extracted structure.
The B and N atoms can be classified into B$_A$, B$_B$ and B$_C$, and N$_A$, N$_B$ and N$_C$.
}
\end{center}
\end{figure*}

\begin{figure*}
\begin{center}
\includegraphics[width=12cm]{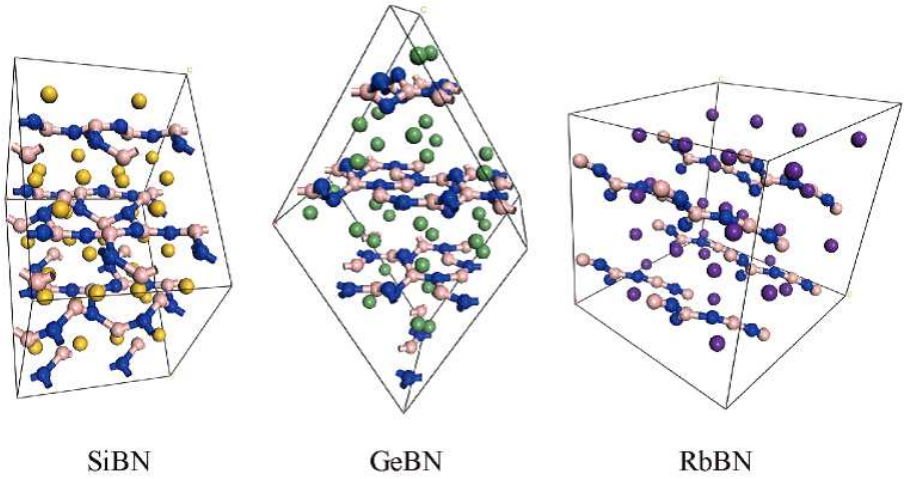}
\caption{\label{fig:K4-SiBN-GeBN-RbBN}
2 $\times$ 2 $\times$ 2 super-cells of fully relaxed SiBN, GeBN and RbBN from $K_{4}$ type crystal structures.
}
\end{center}
\end{figure*}

\begin{figure*}
\begin{center}
\includegraphics[width=12cm]{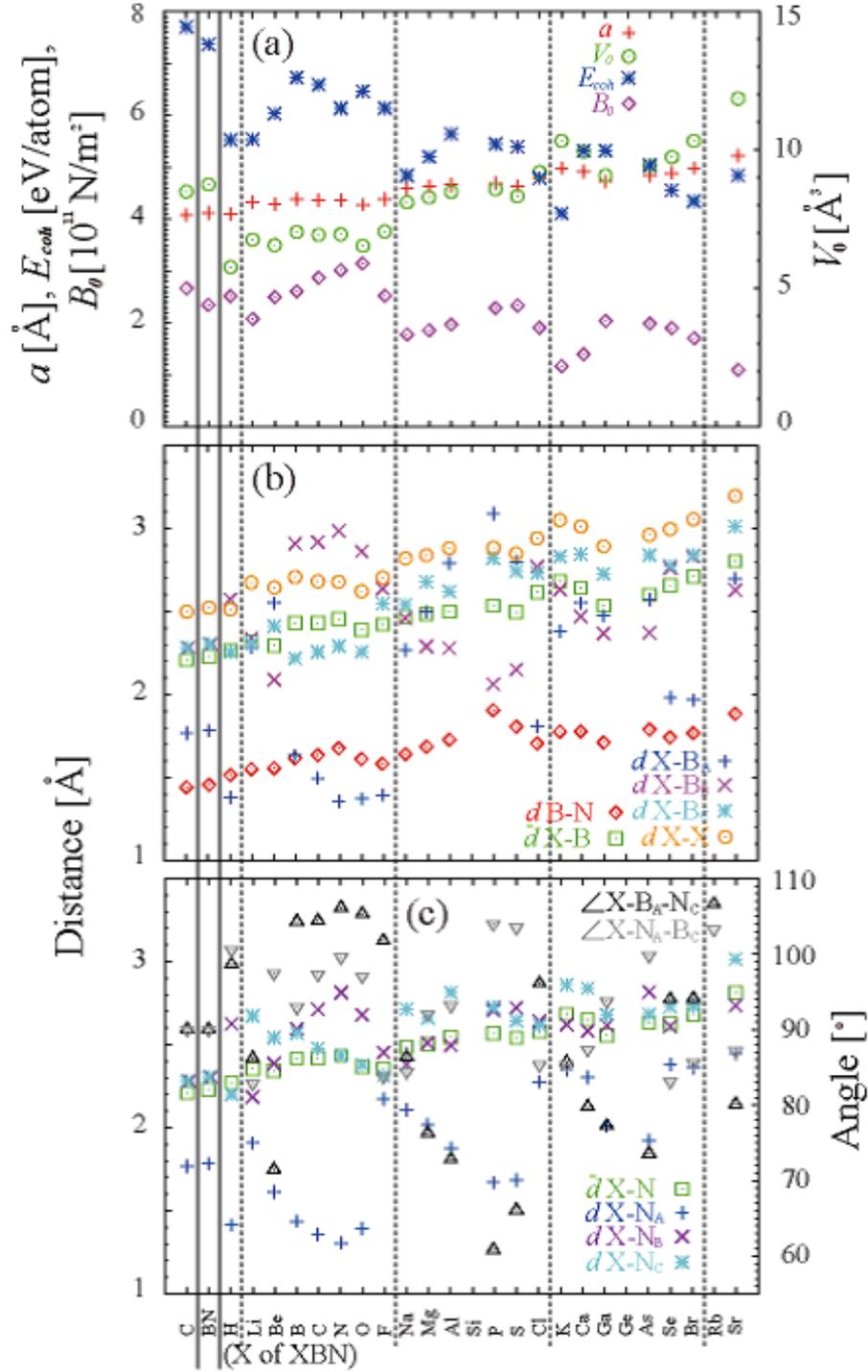}
\caption{\label{fig:K4-C-BN-XBN-structural-detail}
(a) Determined lattice constant($a$), volume at $a$($V_{0}$),
cohesive energy($E_{coh}$), and bulk modulus at $V_{0}$($B_{0}$)
(b) and (c) Nearest-neighbour distances and angles
for the fully optimized B, BN and XBN crystal structures with a $K_4$ type frame within LDA.
Classification of the B and N atoms (B$_A$, B$_B$ and B$_C$, and N$_A$, N$_B$ and N$_C$)
is based on the irreducibility of the crystal symmetry as shown in Fig. 1.
}
\end{center}
\end{figure*}

\begin{figure*}
\begin{center}
\includegraphics[width=10cm]{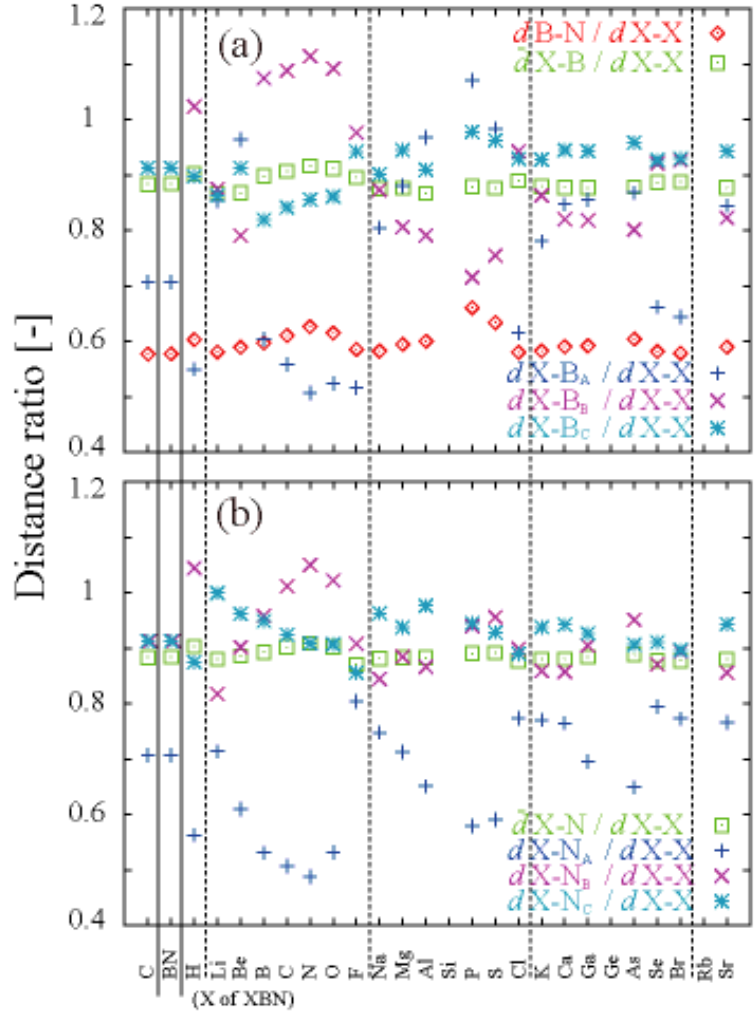}
\caption{\label{fig:K4-C-BN-XBN-d-ratio}
(a) and (b) Distance ratios of the nearest-neighbour distances
for fully optimized BN and XBN crystal structures with $K_4$ type frame within LDA.
Classification of the B and N atoms (B$_A$, B$_B$ and B$_C$, and N$_A$, N$_B$ and N$_C$)
is based on the irreducibility of the crystal symmetry as shown in Fig. 1.
}
\end{center}
\end{figure*}

\begin{figure*}
\begin{center}
\includegraphics[width=12cm]{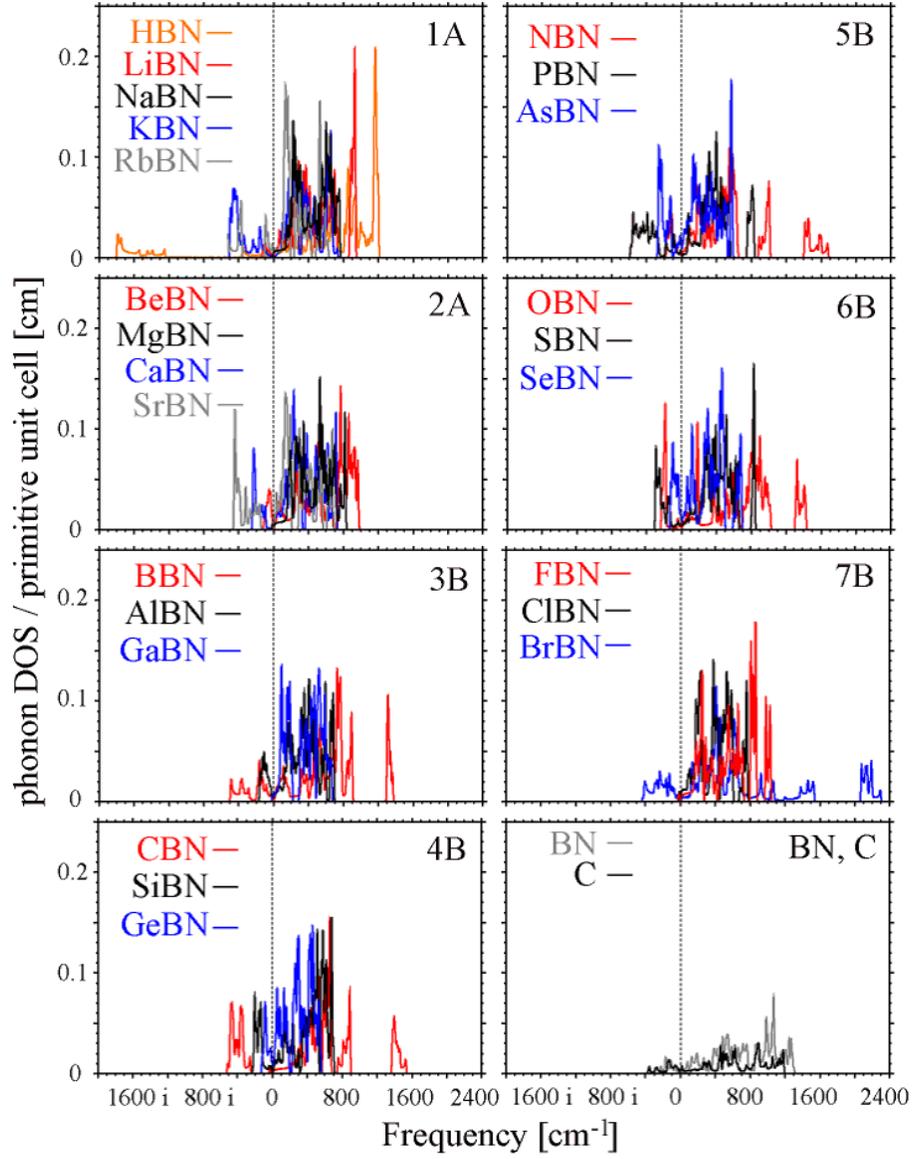}
\caption{\label{fig:K4-XBN-BN-C-sym-phonon-dos}
Phonon density of states(DOS) of XBN (X=H, Li, Be, B, C, N, O, F, Na, Mg, Al, Si, P, S, Cl, K, Ca, Ga, Ge, As, Se, Br, Rb and Sr),
BN and C crystals with $K_{4}$ type frame of BN and C.
The structures are obtained from geometric optimization of the initial structures under a symmetry constraint.
}
\end{center}
\end{figure*}

\begin{figure*}
\begin{center}
\includegraphics[width=12cm]{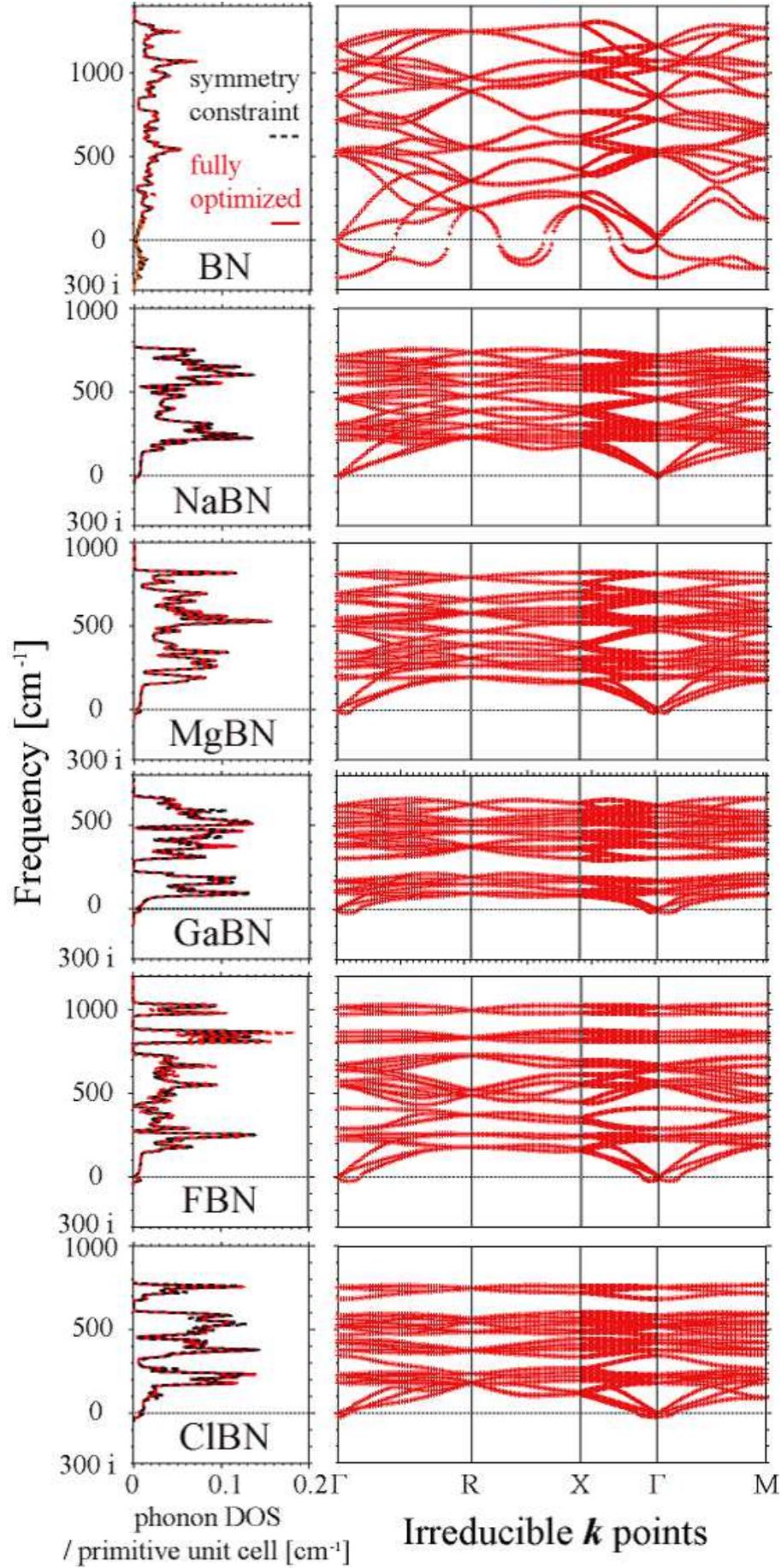}
\caption{\label{fig:K4-BN-NaBN-MgBN-GaBN-FBN-ClBN-sym-non-sym-phonon-dos-band}
(Left) Phonon DOS of partially and fully optimized structures
of BN, NaBN, MgBN, GaBN, FBN and ClBN with a $K_{4}$ type frame of BN.
(Right) Phonon dispersion relationship for the fully optimized structures.
}
\end{center}
\end{figure*}

\begin{figure*}
\begin{center}
\includegraphics[width=17cm]{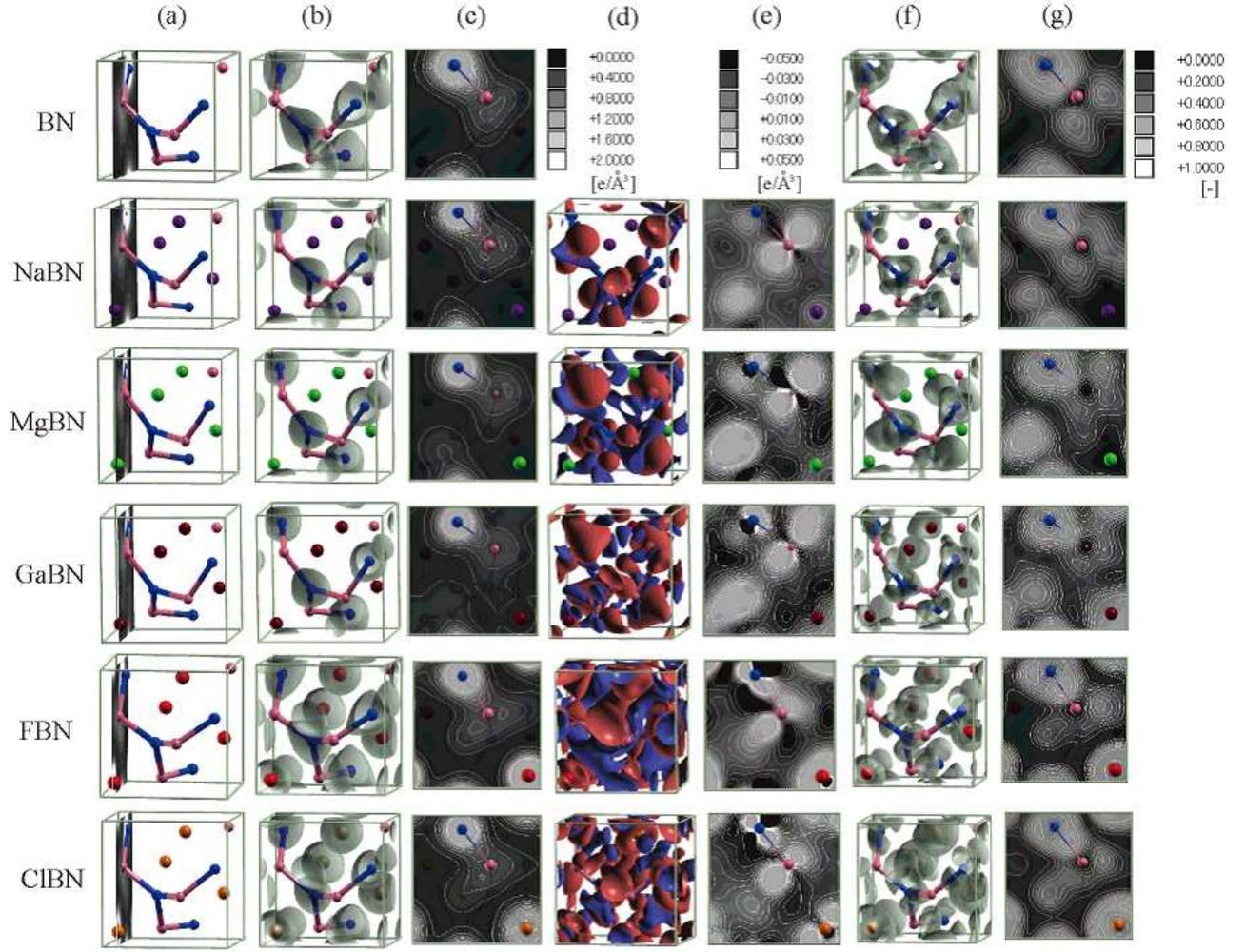}
\caption{\label{fig:K4-BN-NaBN-MgBN-GaBN-FBN-ClBN-chg-diffchg-elf}
(a) Primitive unit cells for fully optimized BN, NaBN, MgBN, GaBN, FBN and ClBN crystals with a $K_4$ type BN frame.
The planes are selected to show the distribution of the valence charge density,
differences in the valence charge density of the crystals (ex. $\rho_{NaBN}(\textbf{r})$) and their separated components:
the induced impurities (ex. $\rho_{Na}(\textbf{r})$) and the frame (ex. $\rho_{BN}(\textbf{r})$),
and the electronic localization function (ELF).
Isosurfaces (b) and contours (c) of the valence charge density (1.0 $e$/\AA $^3$) for BN, NaBN, MgBN, GaBN, FBN and ClBN.
The isosurfaces (d) and contours (e) of the differences in valence charge density
(+(-)0.05$e$/\AA $^3$ coloured red(blue)).
Isosurfaces (f) and contours (g) of the ELF (0.8 [-]).
In (c) and (e), the high and low populations are neglected for comparison at the same density scale.
{XCrySDen}\cite{Kokalj} was used for the visualization.
}
\end{center}
\end{figure*}

\begin{figure*}
\begin{center}
\includegraphics[width=12cm]{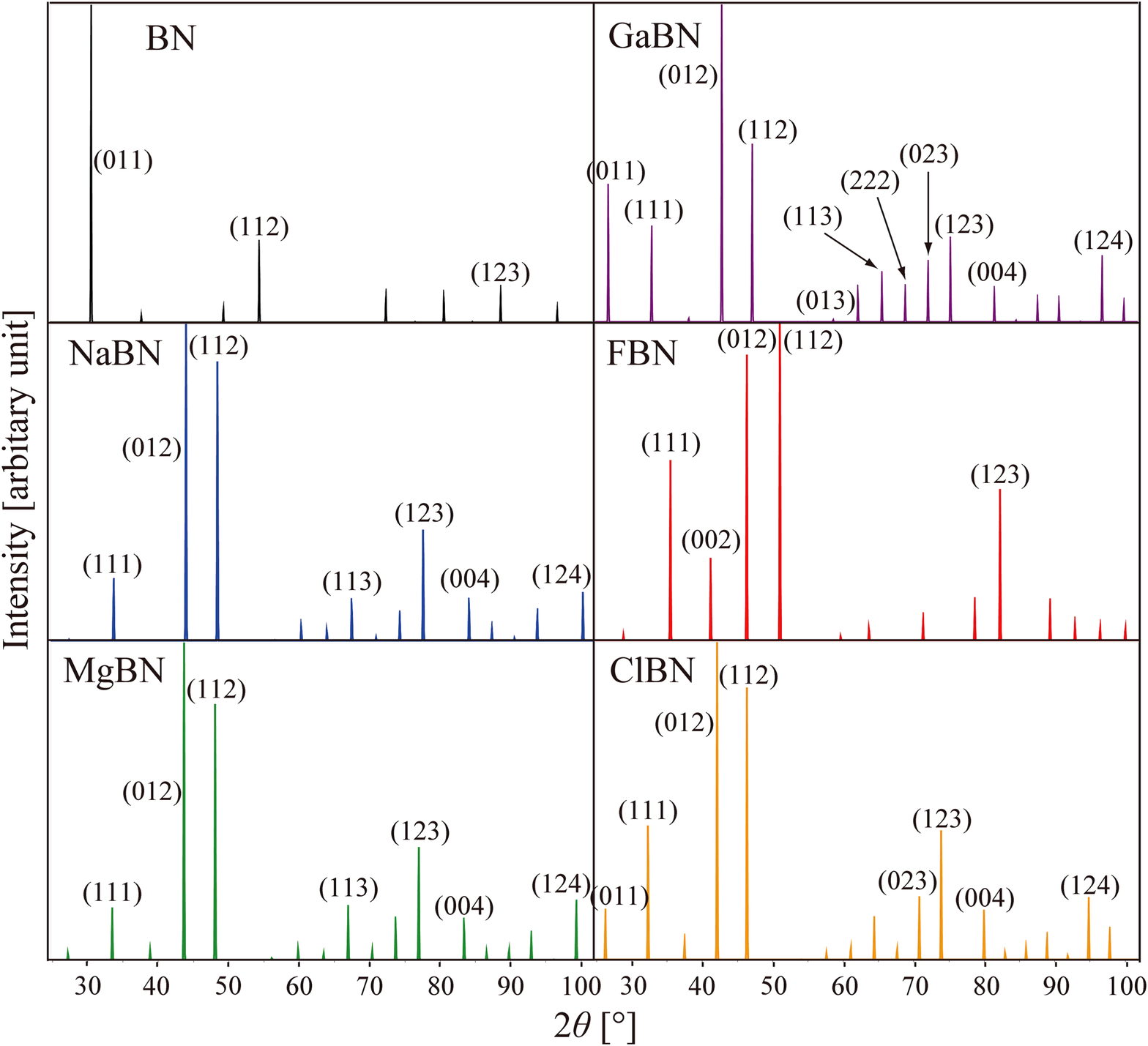}
\caption{\label{fig:K4-BN-NaBN-MgBN-GaBN-FBN-ClBN-XRD}
XRD patterns for the fully optimized structures of BN, NaBN, MgBN, GaBN, FBN and ClBN with $K_{4}$ type BN frames.
}
\end{center}
\end{figure*}

\begin{figure*}
\begin{center}
\includegraphics[width=12cm]{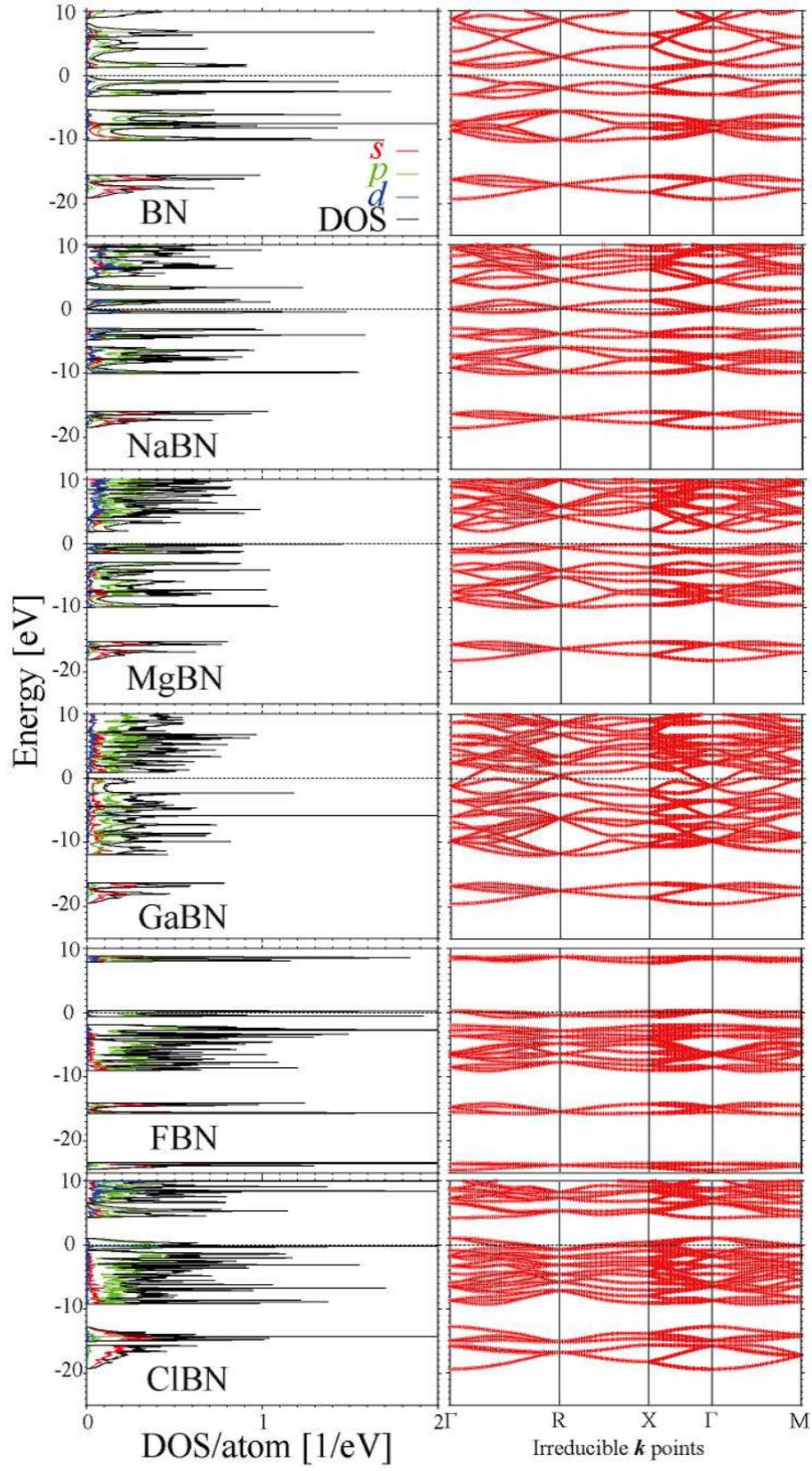}
\caption{\label{fig:K4-BN-NaBN-MgBN-GaBN-FBN-ClBN-KS-dos-band}
Electron density of states and the dispersion relationship for the fully optimized primitive unit cells
of BN, NaBN, MgBN, GaBN, FBN and ClBN with $K_{4}$ type BN frames.
}
\end{center}
\end{figure*}

\begin{table*}
\begin{center}
\caption{\label{tab:expt}
Determined $a$ (lattice constant), $V_{0}$ (volume at $a$),
$E_{coh}$ (cohesive energy) and $B_{0}$ (bulk modulus at $V_{0}$)
for the fully optimized crystal structures within LDA.
}
\begin{ruledtabular}
\begin{tabular}{cccccc}
Component&$a$ [\AA\ ]&$V_0$ [\AA\ $^{3}$/atom]&$E_{coh}$ [eV/atom] & $B_0$ [10$^{11}$ $\frac{N}{m^2}$]\\
\hline
\hline
C\cite{IKNSKA-PRL-2009}&4.082 &8.502 &7.702 &2.67 \\
\hline
\hline
BN&4.121 &8.762 &7.370 &2.35 \\
\hline
\hline
HBN&4.106 &5.767 &5.532 &2.52 \\
\hline
\hline
LiBN&4.330 &6.766 &5.539 &2.08 \\
\hline
BeBN&4.287 &6.565 &6.036 &2.50 \\
\hline
BBN&4.388 &7.042 &6.736 &2.61 \\
\hline
CBN&4.366 &6.936 &6.591 &2.87 \\
\hline
NBN&4.372 &6.962 &6.139 &3.02 \\
\hline
OBN&4.281 &6.539 &6.468 &3.15 \\
\hline
FBN&4.384 &7.053 &6.144 &2.53 \\
\hline
\hline
NaBN&4.596 &8.109 &4.843 &1.78 \\
\hline
MgBN&4.631 &8.291 &5.200 &1.86 \\
\hline
AlBN&4.671 &8.491 &5.645 &1.97 \\
\hline
SiBN (Not $K_{4}$ type)& &9.515 &6.452 &2.06 \\
\hline
PBN&4.687 &8.580 &5.447 &2.29 \\
\hline
SBN&4.642 &8.336 &5.394 &2.34 \\
\hline
ClBN&4.794 &9.202 &4.798 &1.91 \\
\hline
\hline
KBN&4.986 &10.327 &4.114 &1.17 \\
\hline
CaBN&4.924 &9.946 &5.317 &1.40 \\
\hline
GaBN&4.722 &9.087 &5.326 &2.04 \\
\hline
GeBN (Not $K_{4}$ type)& &10.916 &6.213 &1.73 \\
\hline
AsBN&4.840 &9.448 &5.041 &1.99 \\
\hline
SeBN&4.891 &9.753 &4.559 &1.90 \\
\hline
BrBN&4.987 &10.336 &4.346 &1.71 \\
\hline
\hline
RbBN (Not $K_{4}$ type)& &20.057 &5.103 &0.65 \\
\hline
SrBN&5.220 &11.855 &4.844 &1.10 \\
\end{tabular}
\end{ruledtabular}
\end{center}
\end{table*}

\begin{table*}
\begin{center}
\caption{\label{tab:d-angle-detail-1}
Determined nearest-neighbour distances and angles
for the fully optimized C, BN and XBN crystal structures with $K_4$ type frame within LDA.
Classification of the B and N (C) atoms (B$_A$, B$_B$ and B$_C$, and N$_A$, N$_B$ and N$_C$ (C$_A$, C$_B$ and C$_C$))
is based on the irreducibility of the crystal symmetry, as shown in Fig. 1.
For BN (C), X means the atomic position intermediate to the nearest B$_A$ and N$_{A}$ (C$_A$) atoms.
}
\begin{ruledtabular}
\begin{tabular}{ccccccc}
Component&	$d$ B-N [\AA\ ]&$\overline{d}$ X-B [\AA\ ]&$d$ X-B$_A$ [\AA\ ]&	$d$ X-B$_B$ [\AA\ ]& $d$ X-B$_C$ [\AA\ ]& $d$ X-X [\AA\ ]\\
\hline
\hline
C& 1.443& 2.209& 1.768& 2.282& 2.282& 2.500\\
\hline
\hline
BN& 1.457& 2.230& 1.784& 2.304& 2.304& 2.523\\
\hline
\hline
HBN& 1.516& 2.267& 1.381& 2.574& 2.256& 2.514\\
\hline
LiBN& 1.553& 2.318& 2.282& 2.341& 2.308& 2.675\\
\hline
BeBN& 1.559& 2.295& 2.551& 2.091& 2.413& 2.645\\
\hline
BBN& 1.617& 2.431& 1.635& 2.909& 2.218& 2.708\\
\hline
CBN& 1.637& 2.431& 1.497& 2.918& 2.256& 2.682\\
\hline
NBN& 1.678& 2.455& 1.357& 2.985& 2.292& 2.679\\
\hline
OBN& 1.612& 2.391& 1.373& 2.863& 2.258& 2.622\\
\hline
FBN& 1.584& 2.423& 1.396& 2.640& 2.548& 2.705\\
\hline
\hline
NaBN& 1.642& 2.469& 2.269& 2.461& 2.543& 2.820\\
\hline
MgBN& 1.687& 2.486& 2.498& 2.289& 2.680& 2.838\\
\hline
AlBN& 1.730& 2.500& 2.791& 2.280& 2.622& 2.883\\
\hline
SiBN (Not $K_{4}$ type) & & & & & &\\				
\hline
PBN& 1.906& 2.535& 3.088& 2.064& 2.821& 2.885\\
\hline
SBN& 1.806& 2.496& 2.799& 2.150& 2.742& 2.849\\
\hline
ClBN& 1.706& 2.617& 1.811& 2.771& 2.732& 2.941\\
\hline
\hline
KBN& 1.778& 2.682& 2.381& 2.634& 2.831& 3.051\\
\hline
CaBN& 1.779& 2.643& 2.552& 2.471& 2.846& 3.013\\
\hline
GaBN& 1.712& 2.537& 2.474& 2.367& 2.728& 2.893\\
\hline
GeBN (Not $K_{4}$ type)& & & & & &\\						
\hline
AsBN& 1.789& 2.601& 2.571& 2.373& 2.839& 2.962\\
\hline
SeBN& 1.742& 2.657& 1.984& 2.761& 2.777& 2.997\\
\hline
BrBN& 1.769& 2.712& 1.968& 2.833& 2.840& 3.056\\
\hline
\hline
RbBN (Not $K_{4}$ type)& & & & & &\\					
\hline
SrBN& 1.885& 2.804& 2.699& 2.629& 3.013& 3.196\\
\end{tabular}
\end{ruledtabular}
\end{center}
\end{table*}

\begin{table*}
\begin{center}
\caption{\label{tab:d-angle-detail-2}
Determined nearest-neighbour distances and angles
for the fully optimized C, BN and XBN crystal structures with $K_4$ type frame within LDA.
Classification of the B and N atoms (B$_A$, B$_B$ and B$_C$, and N$_A$, N$_B$ and N$_C$)
is based on the irreducibility of the crystal symmetry as shown in Fig. 1.
}
\begin{ruledtabular}
\begin{tabular}{cccccccc}
Component&	$\overline{d}$ X-N [\AA\ ]&$d$ X-N$_A$ [\AA\ ]&$d$ X-N$_B$ [\AA\ ]&	$d$ X-N$_C$ [\AA\ ]& $\angle$ X-B$_A$-N$_C$ [$^{\circ}$]& $\angle$ X-N$_A$-B$_C$ [$^{\circ}$]& $\angle$ B-N-B-N [$^{\circ}$]\\
\hline
\hline
C\cite{IKNSKA-PRL-2009}& 2.209& 1.768& 2.282& 2.282& 90.000& 90.000& 70.529\\
\hline
\hline
BN& 2.230& 1.784& 2.304& 2.304& 90.000& 90.000& 70.529\\
\hline
\hline
HBN& 2.270& 1.415& 2.625& 2.199& 98.641& 100.559& 71.233\\
\hline
LiBN& 2.355& 1.909& 2.186& 2.673& 86.183& 82.908& 75.318\\
\hline
BeBN& 2.342& 1.612& 2.385& 2.543& 71.439& 97.469& 30.344\\
\hline
BBN& 2.418& 1.437& 2.594& 2.569& 104.335& 92.939& 82.667\\
\hline
CBN& 2.419& 1.358& 2.713& 2.478& 104.418& 97.307& 76.571 \\
\hline
NBN& 2.435& 1.306& 2.813& 2.434& 106.143& 99.607& 71.797\\
\hline
OBN& 2.366& 1.394& 2.679& 2.378& 105.311& 97.084& 75.843\\
\hline
FBN& 2.355& 2.174& 2.455& 2.315& 101.720& 83.780& 99.889\\
\hline
\hline
NaBN& 2.485& 2.108& 2.381& 2.715& 86.340& 84.430& 72.238\\
\hline
MgBN& 2.505& 2.021& 2.511& 2.661& 76.250& 92.110& 44.435\\
\hline
AlBN& 2.545& 1.876& 2.497& 2.815& 72.778& 93.194& 37.934\\
\hline
SiBN (Not $K_{4}$ type)& & & & & &\\
\hline
PBN& 2.568& 1.670& 2.709& 2.726& 60.840& 103.998& 12.373\\
\hline
SBN& 2.541& 1.684& 2.722& 2.645& 66.087& 103.532& 17.301\\
\hline
ClBN& 2.579& 2.276& 2.643& 2.617& 96.090& 85.350& 89.292\\
\hline
\hline
KBN& 2.685& 2.351& 2.620& 2.861& 85.652& 85.379& 70.497\\
\hline
CaBN& 2.652& 2.304& 2.582& 2.839& 79.760& 87.374& 57.097\\
\hline
GaBN& 2.558& 2.013& 2.617& 2.680& 77.240& 93.780& 43.431\\
\hline
GeBN (Not $K_{4}$ type)& & & & & &\\
\hline
AsBN& 2.632& 1.922& 2.817& 2.684& 73.552& 99.772& 29.923\\
\hline
SeBN& 2.629& 2.38& 2.610& 2.730& 93.992& 83.161& 89.371\\
\hline
BrBN& 2.682& 2.361& 2.733& 2.739& 94.082& 85.632& 85.293\\
\hline
\hline
RbBN (Not $K_{4}$ type)& & & & & &\\
\hline
SrBN& 2.814& 2.446& 2.735& 3.015& 80.107& 87.151& 58.018\\
\end{tabular}
\end{ruledtabular}
\end{center}
\end{table*}

\begin{figure*}
\begin{center}
\includegraphics[width=10cm]{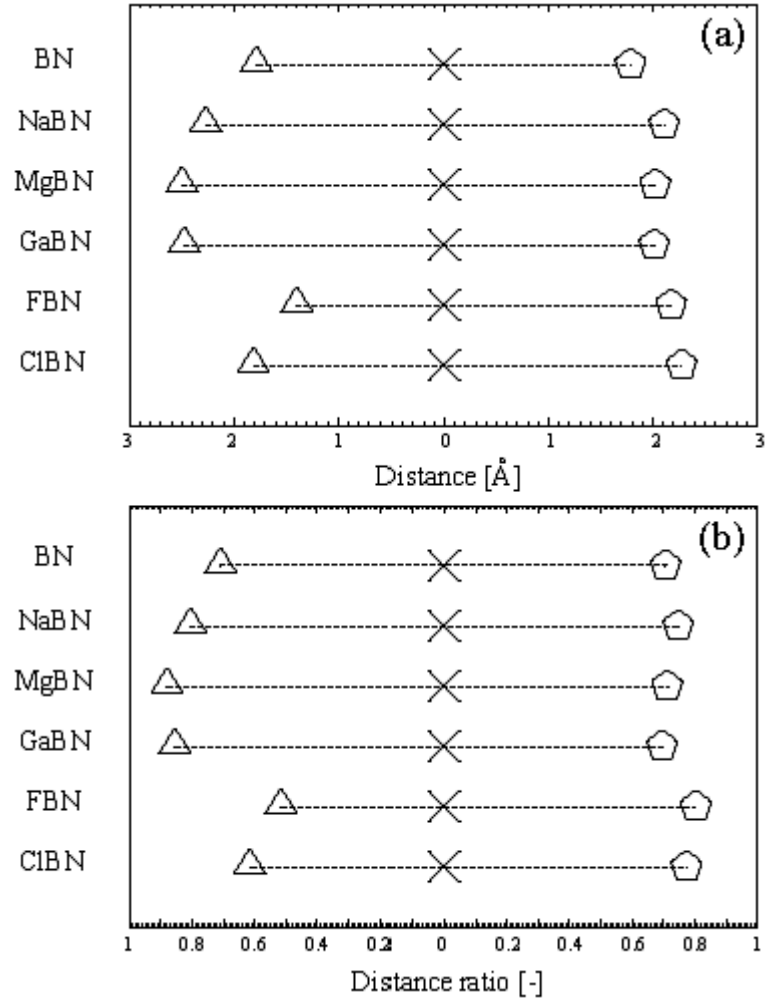}
\caption{\label{fig:K4-BN-NaBN-MgBN-GeBN-FBN-ClBN-d-d-ratio-BA-X-NA}
(a) and (b) Distances and ratios of the nearest-neighbour distances between B$_A$ and X (triangle), and X and N$_A$ (pentagon)
for fully optimized BN, NaBN, MgBN, GaBN, FBN and ClBN crystal structures with $K_4$ type frame within LDA.
Classification of the B and N atoms (B$_A$ and N$_A$)
is based on the irreducibility of the crystal symmetry as shown in Fig. 1.
}
\end{center}
\end{figure*}

\end{document}